\documentclass{ws-ijgmmp}

\begin{document}

\markboth{K.F.Dialektopoulos, S.Capozziello}
{Noether Symmetry Approach}

\catchline{}{}{}{}{}
%
%%%%%%%%%%%%%%%%%%%%%%%%%%%%%%%%%%%%%%%%%%%%%%%%%%%%%%%%%%%%%%%%%%%%

\title{Noether Symmetries as a geometric criterion to select theories of gravity}

\author{Konstantinos F.	Dialektopoulos}
\address{
Dipartimento di Fisica ``E. Pancini", Universit\'a di Napoli
	\textquotedblleft{Federico II}\textquotedblright, Napoli, Italy,\\
INFN Sez. di Napoli, Compl. Univ. di Monte S. Angelo, Edificio G, Via Cinthia, I-80126, Napoli, Italy,\\
\email{dialektopoulos@na.infn.it}
}

\author{Salvatore Capozziello}
\address{
Dipartimento di Fisica ``E. Pancini", Universit\'a di Napoli
	\textquotedblleft{Federico II}\textquotedblright, Napoli, Italy,\\
INFN Sez. di Napoli, Compl. Univ. di Monte S. Angelo, Edificio G, Via Cinthia, I-80126, Napoli, Italy,\\
Gran Sasso Science Institute, Via F. Crispi 7, I-67100, L' Aquila, Italy,\\
Tomsk State Pedagogical University, ul. Kievskaya, 60, 634061 Tomsk, Russia.\\
\email{capozziello@na.infn.it}
}

\maketitle

\begin{history}
\received{(Day Month Year)}
\revised{(Day Month Year)}
\end{history}

\begin{abstract}
We review  the {\it Noether Symmetry Approach} as a geometric criterion to select theories of gravity. Specifically, we deal with Noether Symmetries to solve the  field equations of given  gravity theories. The method allows to  find out exact solutions, but also to constrain arbitrary functions in the action. Specific cosmological models are taken into account.
\end{abstract}

\date{\today}

\section{Introduction}

Even though General Relativity (GR) and the $\Lambda$CDM cosmological model match very well current observations, they present some shortcomings both at cosmological and  astrophysical scales. The discrepancy between the observed value for the cosmological constant with the theoretically calculated vacuum energy of gravitational field  is maybe the biggest open problem in modern physics. But it is not the only one; the inability of experiments or observations to find a convincing particle candidate for dark matter, the existence of singularities in the theory, as well as the inefficiency to find a quantum description of gravitational interactions, made the scientific community pursue for alternative  or extensions  to GR.

Introducing extra fields, higher order derivatives, new degrees of freedom in the form of invariants (e.g. $R_{\mu\nu}R^{\mu\nu},\,R_{\mu\nu\alpha\beta}R^{\mu\nu\alpha\beta},\,\mathcal{G},$ etc.),  torsion, non-metricity and other ingredients are some examples of  modifications adopted to improve the gravitational action, studied in the literature the last decades. 

Some  attempts rely  on adopting arbitrary functions, such as $f(R),\,f(T),\,f(\mathcal{G}),$ etc. where $R$, $T$, $\mathcal{G}$ are curvature, torsion, Gauss-Bonnet scalar invariants respectively. Such theories have to be confronted with data in order to be constrained and then give rise to  physically reliable models. However, apart from phenomenology,  they can also be theoretically constrained and, searching for symmetries, is a straightforward way to do so. Specifically, Noether symmetries can be used as a geometric criterion to choose among modified theories of gravity because the presence of  symmetries identifies  conserved quantities that, in many cases, have also a physical meaning. It is possible to use this theoretical constraint as a sort of selection criterion based on the geometric symmetries of  spacetime \cite{jamal}. This can be obtained by expressing the Lie/Noether symmetry conditions of second order differential equations of the form
\begin{equation}
\ddot{x}^i + \Gamma ^i{}_{jk}\dot{x}^j\dot{x}^k = F^i\,,
\end{equation}
in terms of collineations (i.e. symmetries) of the metric. When this is done, one can use the properties of collineations in differential geometry to find general solutions of the Lie/Noether symmetry problem \cite{Basilakos:2011rx}.

The {\it Noether Symmetry Approach} is outlined in   \cite{Capozziello:1996bi} and in  \cite{Basilakos:2011rx}. Applications to scalar-tensor cosmologies are reported in \cite{Dimakis:2017zdu,Dimakis:2017kwx,Giacomini:2017yuk,Paliathanasis:2014rja}. The most general scalar-tensor theory,  giving second order field equations, the so-called Horndeski gravity, is discussed in \cite{Capozziello:2018gms}. $f(R)$ theories are studied in \cite{Capozziello:2008ch,Paliathanasis:2011jq}, teleparallel gravity and its modifications are discussed in  \cite{Basilakos:2013rua,Bahamonde:2016grb,Capozziello:2016eaz} as well as a class of non-local theories in \cite{Bahamonde:2017sdo}. Apart from cosmology, the method has also been used in spherically and axially symmetric space-time to find exact solutions  \cite{Paliathanasis:2011jq,Capozziello:2009jg,Capozziello:2012iea,Paliathanasis:2014iva,Capozziello:2007wc}.

As a final comment, it is worth  noticing that, apart from the Lie point symmetries\footnote{Noether point symmetries are a subclass of Lie symmetries applied to dynamical systems that are described by a point Lagrangian and leave invariant the action integral.}, which are the simplest kind of symmetries, there exist several other types of transformations for searching for  symmetries in differential equations. In particular, in \cite{Hojman}, S. Hojman proposed a conservation theorem where one uses directly the equations of motion, rather than the Lagrangian or the Hamiltonian of a system and, in general, the conserved quantities can be different from  those derived from the  Noether Symmetry Approach \cite{paolella,andronikos}. In addition, there exist higher order symmetries, such as contact symmetries; that is, when the equation of motion are invariant under contact transformations, which are defined as one parameter transformations, in the tangent bundle of the associated dynamical system \cite{Paliathanasis:2015aos}. Another type of symmetries are the Cartan symmetries \cite{Karpathopoulos:2017arc}, which are point transformations with generators in the tangent bundle, that leave the Cartan 1-form invariant. It has been shown \cite{Cartan1,Cartan2}, that the Cartan symmetries for holonomic dynamical systems are equivalent to the generalized Noether symmetries. In cosmology there are many studies in this direction (see, for example \cite{Paliathanasis:2015gga,Dimakis:2016mip,Barrow:2016wiy} and references therein). The interested reader can  consider  the recent review on symmetries in  differential equations \cite{Tsamparlis:2018nyo}. 

In this paper we want to review the Noether Symmetry Approach, i.e. how to use Noether symmetries as a geometric criterion  to constrain alternative theories of gravity. In addition, we shall work out  an example applying   the method to the $f(R,\mathcal{G})$ theory,where $R$ is the Ricci scalar and $\mathcal{G}$ the Gauss-Bonnet invariant. This case is not yet completely studied in  literature \cite{felix,Capozziello:2014ioa}. Cosmological models are selected according to the existence of the  symmetries and exact cosmological solutions are derived.

The paper is organized as follows: In  Sect. \ref{secII}, we introduce  the point transformations, their properties and how they operate on  differential equations. We discuss how they act on variables, functions and their derivatives. The symmetry generators are presented. In Sect. \ref{secIII}, we define the symmetries of differential equations. Specifically, it is possible to  show that, when  the equations remain invariant under these transformations, then their generator is a symmetry of the equations. In more detail, we apply the method   to dynamical systems, described by a Lagrangian, where the symmetries are the {\it Noether symmetries.} At the end of this section, we construct an algorithm on how to find possible symmetries. In Sect. \ref{secIV} we discuss  $f(R,\mathcal{G})$ gravity as an example. We classify models   admitting symmetries in a cosmological minisuperspace, and we use  symmetries to find exact solutions. Such solutions can have a physical meaning. In Sect. \ref{concl}, conclusions are drawn. 

\section{Point Transformations}
\label{secII}

The aim of this paper is to discuss how to  find  symmetries of differential equations and how to use them  to derive analytic solutions. This cannot be done before we introduce the notion of a symmetry, which in turn, leads  to define the point transformations and their generators.

The mapping of points $(x,y)$ into points $(\bar{x},\bar{y})$, where $x$ is the independent and $y$ is the dependent variable, is called \textit{point transformations.} The one parameter point transformations, 
\begin{equation}\label{eq1}
\bar{x} = \bar{x}(x,y,\epsilon)\,\quad \bar{y} = \bar{y}(x,y,\epsilon)\,,
\end{equation}
are a specific class of point transformations, which have the following properties \cite{Stephani}:
\begin{itemize}
\item
they are invertible,
\item
repeated applications yield a transformation of the same family,
\begin{equation}
\bar{\bar{x}} = \bar{\bar{x}}(\bar{x},\bar{y},\bar{\epsilon}) = \bar{\bar{x}}(x,y,\bar{\bar{\epsilon}})\,, 
\end{equation}
for some $\bar{\bar{\epsilon}} = \bar{\bar{\epsilon}}(\epsilon,\bar{\epsilon})$,
\item
the identity is contained for, say, $\epsilon = 0$
\begin{equation}
\bar{x}(x,y,0) = x\,,\quad \bar{y}(x,y,0) = y \,.
\end{equation}
\end{itemize}

Consider now the one-parameter point transformations \eqref{eq1}. If we expand around $\epsilon = 0$ (= the identity), we get
\begin{eqnarray}\label{eq2}
\bar{x}(x,y,\epsilon) &=& x + \epsilon \frac{\partial \bar{x}}{\partial \epsilon}\vert _{\epsilon = 0} + ... = x + \epsilon \xi (x,y) + ... \\ \label{eq3}
\bar{y}(x,y,\epsilon) &=& y + \epsilon \frac{\partial \bar{y}}{\partial \epsilon}\vert _{\epsilon = 0} + ... = y + \eta (x,y) + ... \,.
\end{eqnarray}
The tangent vector 
\begin{equation}\label{eq6}
\mathbf{X} = \xi(x,y) \frac{\partial}{\partial x} + \eta (x,y) \frac{\partial}{\partial y}\,,
\end{equation}
is called the \textit{infinitesimal generator} of the transformation.

Since our goal is to see how differential equations are affected by these transformations, we have first to extend/prolong them to the derivatives. The transformed derivatives are defined as
\begin{eqnarray}\label{eq4}
\bar{y}' &=& \frac{d\bar{y}(x,y,\epsilon)}{d\bar{x}(x,y,\epsilon)} = \frac{y'(\partial \bar{y}/\partial y)+(\partial \bar{y}/\partial x)}{y'(\partial \bar{x}/\partial y)+(\partial \bar{x}/\partial x)} = \bar{y}'(x,y,y',\epsilon)\,,\\ \label{eq5}
\bar{y}'' &=& \frac{d\bar{y}'}{d\bar{x}} = \bar{y}''(x,y,y',y'',\epsilon)\,, etc.\,.
\end{eqnarray}
By Taylor expanding around $\epsilon = 0$, as we did in \eqref{eq2},\eqref{eq3}, and substitute these into \eqref{eq4},\eqref{eq5}, we obtain
\begin{eqnarray}
\bar{y}' &=& y' + \epsilon \left( \frac{d\eta}{dx} - y' \frac{d\xi}{dx}\right) + ... = y' + \epsilon \eta ^{[1]} + ... \,,\\
\bar{y}^{(n)} &=& y^{(n)} + \epsilon \left( \frac{d\eta^{(n-1)}}{dx}-y^{n}\frac{d\xi}{dx}\right) + ... = y^{(n)} + \epsilon \eta^{[n]}\,,
\end{eqnarray}
where 
\begin{equation}
\eta ^{[n]} = \frac{d\eta ^{(n-1)}}{dx}-y^{(n)}\frac{d\xi}{dx} = \frac{d^n}{dx^n}(\eta -y'\xi) + y^{(n+1)}\xi\,,
\end{equation}
is the ${n}^{th}$ prolongation function of $\eta$. Thus, the ${n}^{th}$ prolongation of the generator $\mathbf{X}$ \eqref{eq6} is
\begin{equation}\label{eq8}
\mathbf{X} ^{[n]}=\mathbf{X} + \eta^{[1]}\partial_{y'} + ... + \eta ^{[n]} \partial _{y^{(n)}}\,. 
\end{equation}
Until now, we referred only to one parameter point transformations. However, the procedure followed to define multiparameter point transformations on variables, their derivatives, as well as their generators, is the same. What we find worth mentioning, since we will use it later on, is what happens to the generating vector if, both the dependent and the independent variables, are more than one. 

Suppose that, a differential equation depends on $r$ independent and $s$ dependent variables, that is $\{ x_i: i = 1,...,r\}$ and $\{y^j: j = 1,...,s\}$, where $y = y (x)$. If we consider the following one parameter point transformation \cite{Kumei}
\begin{equation}\label{eq10}
\bar{x}_i = \Xi _i (x,y,\epsilon)  = x_i + \epsilon \xi_i (x,y) +... \quad , \quad \bar{y}^j = H^j (x,y,\epsilon) = y^j + \epsilon \eta^j (x,y) + ...\,,
\end{equation}	 
with $\xi_k$ and $\eta ^j$ being
\begin{equation}\label{eq11}
\xi _k (x,y) = \frac{\partial \Xi _k (x,y,\epsilon)}{\partial \epsilon}|_{\epsilon\rightarrow 0} \quad , \quad \eta ^j (x,y) = \frac{\partial H^j (x,y,\epsilon}{\partial \epsilon}|_{\epsilon\rightarrow 0}\,.
\end{equation}
the generating vector is given by
\begin{equation}
\textbf{X} = \xi ^k (x^i,y^j) \partial_k + \eta ^l (x^i,y^j) \partial_l\,.
\end{equation}
Following the same procedure as before, we will see how the derivatives of the dependent variables are transformed. At first, we note that $y^j _i = \partial y^j/\partial x_i$ and also
\begin{equation}
D_i = \frac{\partial}{\partial x_i}+ y^j _i\frac{\partial}{\partial y^j} + y^j_{ik}\frac{\partial}{\partial y^j_k}+ ...+y^j_{ii_1i_2...i_n}\frac{\partial}{\partial y^j_{i_1i_2...i_n}} + ...\,.
\end{equation}
Now the derivatives take the form
\begin{align}
\bar{y}^j_i = H^j_i(x,y,\partial y, \epsilon) &= y^j_i+\epsilon\eta^{[1]j}_i(x,y,\partial y) + ...\, \\
&\,\,\,\vdots \nonumber \\
\bar{y}^j_{i_1i_2...i_k} = H^j_{i_1i_2...i_k}(x,y,\partial y,...,\partial ^k y,\epsilon) &= y^j_{i_1i_2...i_k}+\epsilon \eta^{[k]j}_{i_1i_2...i_k} (x,y,\partial y,...,\partial ^k y) +... \,,
\end{align}
where $\eta ^{[1]j}_i = D_i \eta ^j - \left(D_i\xi_k \right) y^j_k$ and $\eta^{[k]j}_{i_1i_2...i_k} = D_{i_k} \eta^{[k-1]j}_{i_1i_2...i_{k-1}} - \left(D_{i_k}\xi_l \right) y^j_{i_1i_2...i_{k-1}l}$. Thus the prolongation of the generator of the point tranformations \eqref{eq10} is given by
\begin{equation}
\mathbf{X}^{[n]} = \mathbf{X} + \eta^{[1]j}_i \partial _{y^j_i} + ... + \eta ^{[k]j}_{i_1i_2...i_k} \partial _{y^j_{i_1i_2...i_k}} \,. 
\end{equation}
These tools will be used to seek for symmetries of differential equations.

\section{Symmetries of differential equations}
\label{secIII}

Now we are ready to study the behaviour of differential equation under the action of point transformations. We already mentioned in the Introduction that apart from Lie/Noether symmetries, which are point symmetries, there exist non-point-like symmetries and higher order symmetries (contact, non-local, Cartan), which we do not discuss in this paper. 

A group of point transformations that maps solutions into solutions, i.e. the mapping $\bar{y}(\bar{x})$ of any solution $y(x)$ is again a solution, is called a \textit{symmetry} of the differential equations. Mathematically formulated it is: the differential equation
\begin{equation}\label{eq7}
H(x,y,y',...,y^{(n)}) = 0\,,
\end{equation} 
remains invariant, under the point transformations (or else symmetry) 
\begin{equation}
\bar{x} = \bar{x}(x,y) \,,\quad \bar{y} = \bar{y}(x,y)\,.
\end{equation}

Consider now a one parameter point transformation and re-express the differential equation \eqref{eq7} in the transformed variables, i.e. $H(\bar{x},\bar{y},\bar{y}',...,\bar{y}^{(n)}) =0$. It is easily seen that,
\begin{equation}
\frac{\partial H(\bar{x},\bar{y},\bar{y}',...,\bar{y}^{(n)})}{\partial \epsilon} \vert_{\epsilon = 0} = \lambda H \Rightarrow \mathbf{X}^{[n]}H = \lambda H\,,
\end{equation}
where $\lambda$ are  eigenvalues. 
The converse is also true and this can be seen only by considering the fact that the existence of symmetries is independent of the choice of variables. Thus we end up with the following theorem:

\textit{\textbf{Theorem:} A differential equation, $H = 0$, admits a group of symmetries with generator $\mathbf{X}$, if and only if $\mathbf{X}^{[n]}H=\lambda H$. 
}

\subsection{Noether point symmetries}

A specific class of Lie point symmetries are the so-called Noether symmetries. They are restricted to dynamical systems coming from a Lagrangian. The Lagrangian function $L = L(t,q^i,\dot{q}^i)$\footnote{The index $i$ takes the values  $1,2,...,n$ and denotes the number of dimensions of the configuration space.}, is a function of time t, the generalized coordinates $q^i=q^i(t)$ and their time derivatives $\dot{q}^i(t)$. It contains information about the dynamics of a system. The equations of motion of the system are given by the Euler-Lagrange equations
\begin{equation}
\frac{d}{dt}\left(\frac{\partial L}{\partial \dot{q}^i}\right) - \frac{\partial L}{\partial q^i} = 0 \,.
\end{equation} 
When the point transformations, that are Lie symmetries for a system of differential equations, transform the Lagrangian in such a way that the action integral remains  invariant, they are called Noether symmetries.

As already known from Lagrangian mechanics, Emmy Noether proved that if a Lagrangian admits a symmetry, then this symmetry is associated with a conserved quantity. The most well known examples are 
\begin{itemize}
\item
the conservation of the total energy of a system, when the Lagrangian is time independent, i.e. it is  not affected by transformations of the form $\bar{t} \rightarrow t+ \delta t$,
\item 
the momentum conservation is associated with the translational invariance, i.e. when a Lagrangian has an ignorable variable, then the associated momentum is conserved,
\item 
and also the angular momentum conservation is related  to the rotational symmetry of the Lagrangian, i.e. the orientation of the physical system in space does not affect the Lagrangian.
\end{itemize}
In a more precise mathematical language:

\textit{\textbf{Theorem:} Suppose a dynamical system described by the Lagrangian $L = L (t,q^i,\dot{q}^i)$. The generator of the infinitesimal point transformations
\begin{align}\label{eq9}
\bar{t} &= t + \epsilon \xi (t,q^i) + ... \,,\\ \label{eq12}
\bar{q}^i &= q^i + \epsilon \eta ^i (t,q^j) + ...\,, 
\end{align}
is 
\begin{equation}\label{eq14}
\mathbf{X} = \xi (t,q^i) \frac{\partial}{\partial t} + \eta ^i \frac{\partial}{\partial q^i}\,,
\end{equation}
while its first prolongation, according to \eqref{eq8}, is given by 
\begin{equation}\label{eq15}
\mathbf{X}^{[1]} = \mathbf{X} + \eta ^{i,[1]}\frac{\partial}{\partial \dot{q}^i}\,.
\end{equation}
If there exists a function $g(t,q^i)$ such that 
\begin{equation}\label{eq13}
\mathbf{X}^{[1]}L + L\frac{d\xi (t,q^i)}{dt} = \frac{dg(t,q^i)}{dt}\,,
\end{equation}
then the Euler Lagrange equations remain invariant under the action of the transformations \eqref{eq9},\eqref{eq12} and $\mathbf{X}$ is the Noether symmetry vector of the system.}

\textit{The associated first integral of motion is given by the function
\begin{equation}\label{eq16}
\phi (t, q^i,\dot{q}^i) = \xi \left(\dot{q}^i \frac{\partial L}{\partial \dot{q}^i}- L\right) - \eta^i \frac{\partial L}{\partial q^i} + g \,.
\end{equation}
}
This is known as the \textit{Noether's second theorem}. In general, Noehter symmetries are valid also for non-point transformations \cite{Noetheroriginal}.

\subsubsection{Invariant functions}
The condition \eqref{eq13} is equivalent to the following Lagrange system
\begin{equation}\label{eq36}
\frac{dt}{\xi} = \frac{dq^i}{\eta^i} = \frac{d\dot{q}^i}{\eta^{[1],i}}.
\end{equation}
The Lagrangian in physical systems  contains up to first order derivatives in the canonical variables, yielding up to second order differential equations. In more general systems with $n$ order differential equations, the above system can be defined with fractions of $n^{th}$ order derivatives of the canonical variables over $n^{th}$ prolongations of the generator coordinates.

The above Lagrange system \eqref{eq36} can give us the zero and first order invariants ($n^{th}$ order in general) respectively
\begin{equation}\label{eq37}
W^{[0]}(t,q^i) = \frac{dt}{\xi} - \frac{dq^i}{\eta^i}\quad \text{and} \quad W^{[1]}(t,q^i,\dot{q}^i) = \frac{dt}{\xi} - \frac{dq^i}{\eta^i} - \frac{d\dot{q}^i}{\eta^{[1]i}}.
\end{equation}    
By using these invariants, we can reduce the order of the Euler-Lagrange equations and thus solve them in an easier way.

\subsection{Finding symmetries}
In the last part of this section, let us see how to find symmetries. We give here a general description of the procedure. In the next section, we will work out a specific example.

In order to determine Noether symmetries, what we need  is to find the coefficients of the generator $\mathbf{X}$, $\xi(t,q)$ and $\eta^i(t,q)$, such that the symmetry condition \eqref{eq13} is satisfied. So, if we have a dynamical system described by a Lagrangian $L = L(t,q^i,\dot{q}^i)$ then:
\begin{enumerate}
\item
We write an \textit{ansatz} for the generator of the form \eqref{eq14} defined on the configuration space.
\item
We expand the symmetry condition \eqref{eq13} to obtain a polynomial depending on $\xi(t,q)$, $\eta^i(t,q)$ and products of the generalized velocities, i.e. $(\dot{q}^a \dot{q}^b...)$. 
\item
Since the unknown coefficients $\xi,\,\eta$ depend only on $(t,q)$, in order for the polynomial to vanish, the coefficients of the products $(\dot{q}^a \dot{q}^b...)$ have to vanish. Thus we end up with a set of partial differential equations for $\xi$ and $\eta$, which, most of the times, can be solved in a straightforward way.
\item 
Once we calculate the generating vector $\mathbf{X}$, we can easily find the first integral $\phi$ from \eqref{eq16}, and thus obtain a better insight into the physical meaning of these integrals.
\item 
Finally, from the generator we can construct the associated Lagrange system, find the zero and first order invariants and reduce the order of the Euler-Lagrange equations.
\item
Depending on the number of symmetries, one can achieve the complete integrability of the dynamical system.
\end{enumerate}

\section{The case of  $f(R,\mathcal{G})$ gravity}
\label{secIV}

Let us now take into account a specific example on how to find out Noether symmetries in modified theories of gravity. Specifically, if a theory of gravity is given by an arbitrary function, e.g. $f(R)$ gravity\footnote{The Einstein-Hilbert action is given by $\mathcal{S}\sim\int d^4x \sqrt{-g} R$, where $R$ is the Ricci scalar. We can extend the action to a more general form, by substituting $R$ with and arbitrary function  $f(R)$ in order to see if we can obtain  physically interesting solutions \cite{Capozziello:2011et,Sotiriou:2008rp,Nojiri:2003ft}.}, adopting the  above procedure, one can  find the  functional form fixed by the existence of  Noether point symmetries.

Let us consider a generalization of $f(R)$ gravity, that is a  theory of the form $f(R,\mathcal{G})$, where $R$ is the Ricci scalar and $\mathcal{G}$ is the Gauss-Bonnet topological invariant  given by
\begin{equation}
\mathcal{G} = R^2 - 4 R_{\mu\nu}R^{\mu\nu} + R_{\alpha\beta\mu\nu}R^{\alpha\beta\mu\nu}\,,
\end{equation}
with $R_{\mu\nu}$ and $R_{\alpha\beta \mu\nu}$ being the Ricci and Riemann tensors respectively. Such a theory is very well motivated for two reasons: firstly, it exhausts all the possible curvature budget by combining $R$ and $\mathcal{G}$; apart from that, Gauss-Bonnet terms arise naturally when trying to renormalize quantum field theories, as well as it contributes to the trace anomaly, too \cite{Capozziello:2011et}. We remind the reader that, a linear term in $\mathcal{G}$, appearing in  the gravitational action (in four dimensions) do not contribute to the field equations being  a topological invariant \cite{Birrell}.

The action of the theory reads
\begin{equation}\label{eq19}
\mathcal{S} = \frac{1}{2\kappa}\int d^4x \sqrt{-g} f\left( R, \mathcal{G}\right) \,,
\end{equation}
where the constant $\kappa = 8 \pi G_N$, with $G_N$ the Newton constant: The speed of light is assumed to be $c=1$. We want to seek for  forms of the function  $f\left( R, \mathcal{G}\right)$ compatible with the existence of  Noether symmetries and then use these symmetries to find out solutions. We will develop our considerations in a cosmological  minisuperspace considering  a  spatially flat Friedman-Robertson-Walker metric of the form
\begin{equation}\label{eq17}
ds^2 = - dt^2 + a(t)^2 \left( dx^2+dy^2+dz^2\right)\,.
\end{equation}
In order to construct a  point-like canonical Lagrangian, we introduce two Lagrange multipliers as
\begin{equation}
\mathcal{S} = \frac{1}{2\kappa}\int dt \sqrt{-g} \left[ f\left( R, \mathcal{G}\right)  -\lambda_1 \left(R-\bar{R} \right) - \lambda_2 \left(\mathcal{G}-\bar{\mathcal{G}} \right) \right]\,,
\end{equation}
where $\bar{R}$ and $\bar{\mathcal{G}}$ are the Ricci scalar and the Gauss Bonnet invariant expressed in terms of the metric \eqref{eq17},
\begin{equation}\label{eq23}
R = 6\left( \frac{\dot{a}^2}{a^2} + \frac{\ddot{a}}{a}\right) \quad , \quad \mathcal{G} = 24 \frac{\dot{a}^2\ddot{a}}{a^3} \,.
\end{equation} 
The Lagrange multipliers $\lambda_1$ and $\lambda_2$ are given by varying the action with respect to $R$ and $\mathcal{G}$ respectively and thus $\lambda_1 = \partial f/\partial R = f_R$ and $\lambda_2 = \partial f/\partial \mathcal{G} = f_{\mathcal{G}}$. After integrating out two total derivatives, we end up with the following point-like Lagrangian
\begin{equation}\label{eq18}
L = a^3 \left(f - \mathcal{G} f_{\mathcal{G}} - R f_{R} \right)  -6 a f_{R} \dot{a}^2  - 6 a^2 f_{R\mathcal{G}} \dot{a} \dot{\mathcal{G}}  - 6 a^2 f_{RR} \dot{a} \dot{R} - 8 f_{\mathcal{G}\mathcal{G}} \dot{a}^3 \dot{\mathcal{G}}  -8 f_{R\mathcal{G}} \dot{a}^3  \dot{R}\,.    
\end{equation}
The configuration space of \eqref{eq18} is $\mathcal{Q} = \{ a,R,\mathcal{G}\}$ and its tangent space $T\mathcal{Q} = \{ a,\dot{a},R,\dot{R},\mathcal{G},\dot{\mathcal{G}}\}$. Hence  the symmetry generating vector is
\begin{equation}\label{eq22}
\mathbf{X} = \xi (t,a,R,\mathcal{G})\partial_t + \eta_a(t,a,R,\mathcal{G})\partial_a+ \eta_R(t,a,R,\mathcal{G})\partial_R+ \eta_{\mathcal{G}}(t,a,R,\mathcal{G})\partial_{\mathcal{G}} \,.
\end{equation}

\subsubsection*{Noether Symmetries}

Let us now apply the symmetry condition \eqref{eq13} to the point-like Lagrangian \eqref{eq18}.  If symmetries exist, such a condition will fix  the form of vector  $\mathbf{X}$ as well as the form $f(R,\mathcal{G})$. By the above procedure,  we obtain an overdetermined  system of 27 partial differential equations. There are several different cases depending on the forms of $f(R,\mathcal{G})$ as well as on the values of integration constants. We summarize them in what follows;  the function $g$ of the right hand side of equation \eqref{eq13} is assumed constant, unless otherwise stated. All the $c_i$'s constant mentioned below  are integration constants.
There are two different classes of theories; those for which the derivative $f_{R\mathcal{G}}$ vanishes and those for which it does not. 
\begin{enumerate}
\item
For $f_{R\mathcal{G}}\neq  0$ we have the following cases:
\begin{enumerate}
\item
If $f_{RR} \neq 0 $, then 
\begin{itemize}
\item
for $c_1 \neq c_2$, we get
\begin{equation}\label{eq26}
f(R,\mathcal{G}) = R^{\frac{c_1+3 c_2}{2 c_1}} \tilde{f}\left(\frac{\mathcal{G}}{R^2}\right)+\frac{c_4}{3 \left(c_2-c_1\right)}\mathcal{G}\,,
\end{equation}
where $\tilde{f}$ is an arbitrary function of $\mathcal{G}/R^2$. It admits the Noether symmetry  vector
\begin{equation}\label{eq24}
\mathbf{X} = (c_1 t+c_3)\partial _t +c_2 a \partial _a - 2 c_1 R \partial _R - 4 c_1 \mathcal{G} \partial _{\mathcal{G}}\,.
\end{equation}
\item
For $c_1 = c_2$,  the theory 
\begin{equation}\label{eq27}
f(R,\mathcal{G}) = R^2 \tilde{f}(\frac{\mathcal{G}}{R^2})-\frac{ c_4 \mathcal{G}  \ln R}{2 c_1}\,,
\end{equation}
admits the Noether symmetry
\begin{equation}\label{eq25}
\mathbf{X} = (c_1 t+c_3)\partial _t +c_1 a \partial _a - 2 c_1 R \partial _R - 4 c_1 \mathcal{G} \partial _{\mathcal{G}}\,.
\end{equation}
\end{itemize}
\item
If $f_{RR} = 0$, the function $f$ takes the form $f(R,G) = f_1(\mathcal{G}) + R f_2(\mathcal{G})$ and the following cases are obtained:
\begin{itemize}
\item
For $c_1 = 0$, the theory $f(R,G) = f_1(\mathcal{G})+R f_2(\mathcal{G})$ admits only the symmetry $\mathbf{X} = c_3 \partial _t$ and the associated integral is the Hamiltonian $E$\,.
\item
For $c_1 \neq 0 $ we have two cases
\begin{enumerate}
\item
If $c_2 \neq c_1$, then the theory
\begin{equation}\label{eq28}
f(R,\mathcal{G}) =\frac{c_4}{3 \left(c_2-c_1\right)}\mathcal{G}+c_5 \mathcal{G}^{\frac{c_1+3 c_2}{4 c_1}}+ c_6 R \mathcal{G}^{\frac{3 c_2-c_1}{4 c_1}} \,,
\end{equation}
admits the symmetry \eqref{eq24}. 
\item 
If $c_2 = c_1$, the theory
\begin{equation}\label{eq29}
f(R,\mathcal{G}) = -\frac{c_4 \mathcal{G} \ln \mathcal{G}}{4 c_1}+c_5 \mathcal{G} + c_6 \sqrt{\mathcal{G}} R\,,
\end{equation}
admits the symmetry \eqref{eq25}.
\end{enumerate} 
\end{itemize}
\end{enumerate}
\item
The second class of theories are those for which $f_{R\mathcal{G}} = 0$, which means $f(R,\mathcal{G} ) = f_1(R) + f_2 (\mathcal{G})$. For these theories we find that
\begin{enumerate}
\item
If $f_1''(R) \neq 0$ and 
\begin{enumerate}
\item 
$f_2''(\mathcal{G}) \neq 0$, there are three possible cases:
\begin{itemize}
\item
If $c_2 \neq c_1$ and $c_2 \neq - c_1/3$ the theory
\begin{equation}\label{eq30}
f(R,\mathcal{G}) =\frac{c_4 \mathcal{G}}{3 c_2-3 c_1} + c_5 R^{\frac{c_1+3 c_2}{2 c_1}}+c_6 \mathcal{G}^{\frac{c_1+3 c_2}{4 c_1}}\,,
\end{equation}
admits the symmetry vector given by Eq. \eqref{eq24}.
\item
If $c_2 = c_1$ the theory 
\begin{equation}\label{eq31}
f(R,\mathcal{G}) = -\frac{c_4 \mathcal{G} \ln \mathcal{G}}{4 c_1} + c_5 R^2 + c_6 \mathcal{G}\,,
\end{equation}
admits the Noether vector given by Eq. \eqref{eq25}. 
\item
If $c_2 = - c_1/3$, the theory 
\begin{equation}\label{eq32}
f(R,\mathcal{G}) = - \frac{c_4}{4 c_1} \mathcal{G}+\frac{c_5}{2 c_1} \ln \left(\frac{\sqrt{\mathcal{G}}}{R}\right)+c_6 \,,
\end{equation}
admits the symmetry with generator 
\begin{equation}
\mathbf{X} = (c_1 t+c_3) \partial _t - \frac{c_1}{3}a\partial _a - 2 c_1 R\partial _R - 4 c_1 \mathcal{G} \partial _{\mathcal{G}}\,.
\end{equation} 
\end{itemize}
\item
$f_2''(\mathcal{G}) = 0$, then $f$ takes the form $f(R,\mathcal{G}) = f_1(R) + c_1 \mathcal{G}+ c_2,$ and the system reduces to three equations, but it is not solvable for arbitrary $f_1(R)$.
\end{enumerate}
\item
If $f_1''(R) = 0$ then we can write it as $f_1(R) = c_5 R + c_6$. Thus we end up with the following cases 
\begin{enumerate}
\item
For  theories which are General Relativity with $c_5 \neq 0$
\begin{enumerate}
\item
and $f_2''(\mathcal{G}) \neq 0$, there are two different cases, one of which is non-trivial:
\begin{itemize}
\item
when $c_1 \neq 0 $, the theory
\begin{equation}\label{eq33}
f(R,\mathcal{G}) = c_5 R - \frac{c_4}{2 c_1}\mathcal{G} + c_7 \sqrt{\mathcal{G}}\,,
\end{equation}
admits the following symmetry
\begin{equation}
\mathbf{X} = (c_1 t + c_3)\partial _t + \frac{c_1}{3} a \partial _a - 4 c_1 \mathcal{G} \partial _{\mathcal{G}}\,,
\end{equation}
\item
while when $c_1 = 0$, the theory is $f(R,\mathcal{G}) = c_5 R + c_6 + f_2(\mathcal{G})$ and admits only the symmetry $\mathbf{X} = c_3 \partial _t$ with the Hamiltonian as integral of motion.
\end{itemize}
\item
and $f_2''(\mathcal{G}) = 0 \Rightarrow f_2(\mathcal{G}) = c_7 \mathcal{G}$ we obtain the following two non-trivial cases:
\begin{itemize}
\item 
For $c_6 \neq 0$ the theory has the form
\begin{equation}
f(R,\mathcal{G}) = c_5 R + c_6 + c_7 \mathcal{G}\,, 
\end{equation}
and the Noether vector becomes
\begin{align}
\mathbf{X} =&\left[\frac{a}{3} \left(c_3 \sin \left(\frac{t}{c_0}\right)+c_2 \cos \left(\frac{t}{c_0}\right)\right)+\frac{1}{\sqrt{a}}\left( c_4 \sin \left(\frac{t}{2c_0}\right)+c_8 \cos \left(\frac{t}{2c_0}\right)\right)\right] \partial _a+\nonumber \\
&+\left[c_1 -c_0 \left(c_2 \sin \left(\frac{t}{c_0}\right)+c_3 \cos \left(\frac{t}{c_0}\right)\right)\right]\partial_t \,,
\end{align}
where $c_0 = \sqrt{2 c_5/(3c_6)}$ is a redefinition of the integration constants. In addition, while in all the previous cases, the right hand side function $g$ in Eq. \eqref{eq13} was constant, here it becomes non-trivial and takes the form
\begin{align}
g(a,t) = &2 \sqrt{c_5c_6}  \Big[\sqrt{6} a^{3/2} \left(c_8 \sin \left(\frac{t}{2 c_0}\right)-c_4 \cos \left(\frac{t}{2 c_0}\right)\right)+\nonumber \\
&+\sqrt{\frac{2}{3}} a^3 \left(c_2 \sin \left(\frac{t}{c_0}\right)-c_3 \cos \left(\frac{t}{c_0}\right)\right)\Big] + g_0\,.
\end{align}
\item 
For $c_6 = 0$ the theory is the Einstein-Hilbert action plus a topological invariant term (Gauss-Bonnet) $f(R,\mathcal{G})  = c_5 R+ c_7 \mathcal{G}$ which does not contribute to the dynamics in 4-dimensional spaces. In this case, the Noether vector takes the form
\begin{equation}
\mathbf{X} = \left(t \left(c_1 t+c_2\right)+c_3\right)\partial _t+ \left(\frac{1}{3} a \left(2 c_1 t + c_2\right)+\frac{c_4 t+c_8}{\sqrt{a}}\right)\partial _a\,,
\end{equation}
and the function $g$ is again non-trivial
\begin{equation}
g(a) = -8 a^{3/2} c_4 c_5 - \frac{8}{3} a^3 c_1 c_5 + g_0\,.
\end{equation}
\end{itemize}
\end{enumerate}
\item
In the case where the theories do not contain any Ricci scalar in the action, i.e. $c_5 = 0$, the function $f$ takes the form $f(R,\mathcal{G}) = c_6 + f_2(\mathcal{G})$. The only interesting theories in this case, are those for which $f_2''(\mathcal{G}) \neq 0 $, since if $f_2''(\mathcal{G}) = 0 $, we will have only linear terms in $\mathcal{G}$ in the action and the theory is  trivial for the above reasons.  So, for $f_2''(\mathcal{G}) \neq 0 $ and
\begin{itemize}
\item $c_2\neq c_1$,  we have the theory
\begin{equation}\label{eq34}
f(R,\mathcal{G}) = c_7 \mathcal{G}^{\frac{c_1+3 c_2}{4 c_1}}+\frac{c_4 }{3 \left(c_2-c_1\right)}\mathcal{G}\,,
\end{equation}
and its Noether symmetry is given by
\begin{equation}
\mathbf{X} = (t c_1 + c_3) \partial _t + c_2 a \partial_a - 4 c_1 \mathcal{G}  \partial_{\mathcal{G}}\,;
\end{equation}
\item 
for $c_2= c_1$,  the theory takes the form
\begin{equation}\label{eq35}
f(R,\mathcal{G}) = c_7 \mathcal{G} - \frac{c_4 }{4 c_1}\mathcal{G}\ln\mathcal{G}\,,
\end{equation}
and its generator
\begin{equation}
\mathbf{X} = (t c_1 + c_3) \partial _t + c_1 a \partial_a - 4 c_1 \mathcal{G}  \partial_{\mathcal{G}}\,.
\end{equation}
\end{itemize}
\end{enumerate}
\end{enumerate}
\end{enumerate}
It is obvious that, for each of these functions $f$ and its generators, there exists an integral of motion, which we do not report  here  for the sake of simplicity. In conclusion,  we have showed that,  by  Noether theorem,  it possible to select specific models   of a given theory of gravity, in this case  $f(R,\mathcal{G})$ of the action \eqref{eq19}. Models that admit Noether symmetries have   generating vectors and  associated integrals of motion. These symmetries can be used to find analytical solutions as we we see below.

\subsubsection*{Cosmological Solutions}

Finding out  exact solutions is the main issue related to the search for symmetries. In other words, if the symmetries do not reduce  dynamics, they are useless from the point of view of dynamical systems. Here we consider some specific forms of $f(R,\mathcal{G})$, selected above by the existence of the Noether symmetries, and search for  cosmological solutions.
From the generator of  symmetries,  we can calculate the zero-order invariants for each case. 
The Noether vector for several of  above models, i.e.  \eqref{eq26}, \eqref{eq27}, \eqref{eq28}, \eqref{eq29}, \eqref{eq30}, \eqref{eq31}, \eqref{eq32}, \eqref{eq33}, \eqref{eq34}, \eqref{eq35}, has the same form, with different constants of integration. From this, we get the Lagrange system
\begin{equation}
\frac{dt}{c_1 t} = \frac{da}{c_2 a} = -\frac{dR}{2 c_1 R} = -\frac{d\mathcal{G}}{4 c_1 \mathcal{G}}\,,
\end{equation}
and solving for $a(t),\,R(t)$ and $\mathcal{G}(t)$, we get 
\begin{equation}
a(t) = a_0 t^{c_2/c_1}\,,\,\,R(t) = \frac{r_0}{t^2} \,,\,\,\mathcal{G}(t) = \frac{g_0}{t^4}\,,
\end{equation}
where $a_0 ,\,r_0$ and $g_0$ are constants. By substituting these into the Euler-Lagrange equations for $a,\,R$ and $\mathcal{G}$ we can constrain the arbitrary functions $\tilde{f}(\mathcal{G}/R^2)$ in the \eqref{eq26} and \eqref{eq27}, as well as the integration constants $c_i$'s.

Let us study the different models one by one. Consider the model given by \eqref{eq26}. We substitute it into the point-like Lagrangian \eqref{eq18} and write down the Euler-Lagrange equations for the variables of the configuration space. The  equations for $R$ and for $\mathcal{G}$, as expected from the Lagrange multipliers, give the expressions \eqref{eq23}. It is easy to  see that the cosmological scale factor  $a$ has power law solutions of the form $a (t) = a_0 t^p$ for  $c_2 = c_1(3p-1)/3$,  and de-Sitter solutions $a(t) = a_0 e^{H_0 t}$, for  $\tilde{f}(1/6) = 0$. 

In the same way, the model given by \eqref{eq27} admits de Sitter solutions for $c_4 = 0$. Power-law solutions are obtained  for some specific values of $\tilde{f}$. The model \eqref{eq28} admits power-law solutions for $c_2 = c_1 (3p-1)/3$ and de-Sitter solutions for $c_6 = - c_5/\sqrt{6}$. The model \eqref{eq29} gives de-Sitter solutions as soon as $c_4 = 0$ and power-law solutions for
$$
c_6 = \frac{c_4 \sqrt{(p-1) p^3} (p+3)}{3 \sqrt{6} c_1 (p-1) p}\,.
$$
For the model \eqref{eq30} power-law solutions exist for $c_2 = c_1(3p-1)/3$ and de-Sitter solutions for $c_6 = -6^{(3c_2+c_1)/4c_1}c_5$. The model \eqref{eq31} admits de-Sitter solutions only for $c_4 = 0$, but gives power-law solutions for
$$
c_4 = \frac{18c_5 (2p-1)c_1}{p(p+3)}\,.
$$
The model \eqref{eq32} gives de-Sitter solutions for $c_6 = c_5 \ln 6/4c_1$ and power-law solutions for
$$
c_6=-\frac{c_5 \left(\left(4 p^2-6 p+2\right) \ln \left(-\frac{\sqrt{\frac{2}{3}} \sqrt{(p-1) p^3}}{p-2 p^2}\right)+3 p+1\right)}{4 c_1 \left(2 p^2-3 p+1\right)}\,.
$$
The model \eqref{eq33} admits de-Sitter solutions for $c_7 = -\sqrt{6} c_5$ and power-law solutions for $c_7 = \sqrt{6}c_5 \frac{p^3(p-1)}{p(p+1)}$. The model \eqref{eq34} admits only power-law solutions for $c_2 = - c_1/3$ or $c_2 = c_1 (3p-1)/3$. Finally, the model \eqref{eq35} admits only  power-law solutions for $p=-3$ and $p = 4/3$.

\section{Conclusions}
\label{concl}

Symmetries can be considered a general criterion to select physical models. In this paper, we discussed a specific class of symmetries, the so-called  Noether symmetries, which  are Lie symmetries for dynamical systems derived from a  Lagrangian. Specifically, we presented the one-parameter point transformations that maintain \textit{second} order differential equations invariant. It is a purely geometric approach which allowing to  classify  theories of gravity that admits Noether symmetries, gives rise to the possibility to  find exact solutions. 

As  specific example, we presented  $f(R,\mathcal{G})$ gravity. We showed that if the symmetry generator vector exists, it fixes the form of the  $f(R,\mathcal{G})$ function and allows to find out exact cosmological solutions which are physically interesting being Eistein-de Sitter power law solutions or de Sitter exponential solutions. In this sense, the Noether Symmetry Approach is a natural geometric criterion capable of selecting physical model. This statement is particularly relevant in Quantum Cosmology where the existence of symmetries allow to select observable universes according to the prescriptions of the Hartle criterion (see \cite{sergei} for details).

\section*{Acknowledgments}
The Authors  thank the organizers of the workshop: {\it Geometric Foundations of Gravity} (September 2017), held in Tartu, Estonia, for their warm hospitality.
 They are partially  supported by the INFN sezione di Napoli (iniziative specifiche TEONGRAV and QGSKY). This paper is based upon work from COST Action CA 15117 (CANTATA), supported by COST (European Cooperation in Science and Technology).

\end{document}